\documentclass[a4paper,12pt]{article}
\usepackage[left=2.5cm,bottom=3cm,right=2.5cm,top=3cm]{geometry} 
 \usepackage{mathtools}  
\usepackage{makecell}   %
\usepackage{bigints}
\usepackage{overpic,youngtab}
\usepackage{subfigure}
\usepackage{bbm}
\usepackage[latin,english]{babel}
\usepackage{amsmath}
\usepackage{amssymb}
\usepackage{epstopdf}
\usepackage{graphics,psfrag,rotating}
\usepackage{mathabx}
\usepackage{graphicx}
\usepackage{dcolumn}
\usepackage{float}
\usepackage{pdflscape}
\usepackage{array}
\usepackage{booktabs}
\usepackage{amscd} 
\usepackage{mathtools}
\usepackage{fancybox}
\usepackage{fix-cm}
\usepackage[colorlinks=true,citecolor=blue,,linktocpage=true,linkcolor=blue,urlcolor=black]{hyperref}
\usepackage{tikz} 
\usetikzlibrary{matrix} 
\usetikzlibrary{positioning} 
\usetikzlibrary{arrows}
\usepackage{titlesec}
\usepackage{abstract}
\newcolumntype{M}[1]{>{\centering\arraybackslash}m{#1}}
\newcolumntype{N}{@{}m{0pt}@{}}

\def\be{\begin{equation}}
\def\ee{\end{equation}}
\def\bea{\begin{eqnarray}}
\def\eea{\end{eqnarray}}
\def\pd{\partial}
\def\a{\alpha}
\def\b{\beta}
\def\g{\gamma}
\def\d{\delta}
\def\m{\mu}
\def\n{\nu}
\def\t{\tau}

\def\l{\lambda}

\def\r{\rho}

\def\vp{\vec{p}}

\def\bR{\bar{R}}

\def\bp{\bar{\phi}}

\def\bW{\overline{W}}
\def\bn{\bar{\nabla}}
\def\bR{\bar{R}}
\def\bW{\bar{W}}

\def\s{\sigma}

\def\e{\epsilon}
\def\bi{\begin{itemize}}
\def\ei{\end{itemize}}
\def\bg{\bar{g}}

\def\vp{\varphi}
\def\bvp{\bar{\varphi}}
\newcommand{\email}[1]{\href{mailto:#1}{\tt #1}}

\begin{document}

		\vspace*{-1cm}
		\phantom{hep-ph/***} 
		{\flushleft
			{{FTUAM-19-14}}
			\hfill{{ IFT-UAM/CSIC-19-90}}}
		\vskip 1.5cm
		\begin{center}
		{\LARGE\bfseries Scalar Weyl anomalies and the dynamics of the  gravitational field.}\\[3mm]
			\vskip .3cm
		
		\end{center}
		\vskip 0.5  cm
		\begin{center}
			{\large Enrique Alvarez,
			Jesus Anero and Raquel Santos-Garcia  }
			\\
			\vskip .7cm
			{
				Departamento de F\'isica Te\'orica and Instituto de F\'{\i}sica Te\'orica, 
				IFT-UAM/CSIC,\\
				Universidad Aut\'onoma de Madrid, Cantoblanco, 28049, Madrid, Spain\\
				\vskip .1cm

				\vskip .5cm
				\begin{minipage}[l]{.9\textwidth}
					\begin{center} 
						\textit{E-mail:} 
						\email{enrique.alvarez@uam.es,jesusanero@gmail.com, raquel.santosg@uam.es}

					\end{center}
				\end{minipage}
			}
		\end{center}
	\thispagestyle{empty}
	
\begin{abstract}\vspace{-1em}
	\noindent
The generalization  of scale invariance  when  gravitational effects are considered is Weyl invariance, namely, invariance under (global or local)  rescalings of the metric. In this work, we discuss in some details the implications of the fact that the value of the anomaly for the global Weyl invariant coupling of scalar fields to gravity is 
sensitive to the  dynamics (or absence thereof)  of the gravitational field. 
\end{abstract}
\newpage
\tableofcontents
	\thispagestyle{empty}
\flushbottom

\newpage
\setcounter{page}{1}
\section{Introduction}
Shortly after gauge anomalies associated with inconsistencies of chiral fermionic theories in the presence of a {\em background}, non-dynamical, gauge field were first discovered, Gross and Jackiw \cite{Gross} underwent the study of what happens when the gauge field was dynamical. They found that theories that were believed to be renormalizable owing to gauge invariance, lost this property precisely because this same gauge invariance was broken by the anomaly.
\par
 With this in mind, we would like to examine the effects of a dynamical gravitational field on theories that suffer from a Weyl anomaly. This just means, of course, that we are integrating in the  Feynman functional integral over metric field fluctuations, instead of consider the metric as a fixed classical background field. 
 \par
 The situation in our case  is different form the one involving ordinary gauge fields with compact gauge groups, because gravitational theories are not renormalizable to begin with (cf. for example the review \cite{Alvarez}); but Ward identities still do depend on whether gravitation behaves as an ordinary quantum field or not, and we would like to examine precisely in what sense.
\par
The particle nature of the gravitational interaction is grounded in the existence of a spin 2 particle mediating the interaction on a flat background, as first introduced by Fierz and Pauli in \cite{Fierz}. This picture is consistent with General Relativity when an interacting theory for the gravitons is constructed in a consistent way \cite{Deser}. There has always been a current of opinion maintaining that gravitation should not be quantized, because is the only interaction with a clear geometric interpretation. To mention a recent particular approach, the entropic origin of the gravitational interaction, first introduced by Jacobson \cite{Jacobson} and recently popularized by Padmanabhan and Verlinde \cite{Padmanabhan,Verlinde}, may give an alternative explanation which is compatible with Einstein's equation.  No experiment as yet has proved conclusively that the gravitational field has to be quantized;  may be it is a purely classical field after all. One standard argument in favor of that is that it provides  the arena for all other interactions to propagate. This is what we mean by a {\em non-dynamical} gravitational field.
The aim of this work is to highlight the dependence of the anomaly on the  character of the gravitational field and to quantify such effect. 
\par
This computation is unusual; normally anomalies are computed with gauge fields acting as a non-dynamical background. In particular, no many anomalies with quantum gravity effects included are known to us. There is always a trivial modification  because loops of gravitons are missing in the classical background approach. Although this is undoubtly true, the real difference is somewhat subtler, and the our purpose here is to clarify it.
\par

One computation of the anomaly including quantum dynamical gravitational field  was already carried out in \cite{EA} for the non minimal coupling of the scalar field to gravity. We shall examine another example of globally Weyl invariant model with minimal coupling and compare their respective anomalies. 
Although all our computations were done for scalar fields, we do not expect anything particularly new when including fermions.
\par


\par
Throughout  this work we follow the Landau-Lifshitz spacelike conventions, in particular
\be
R^\m_{~\n\r\s}=\partial_\r \Gamma^\m_{\n\s}-\partial_\s \Gamma^\m_{\n\r}+\Gamma^\m_{\l\r}\Gamma^\l_{\n\s}-\Gamma^\m_{\l\s}\Gamma^\l_{\n\r}\, . \ee
The Ricci tensor takes the form
\be R_{\m\n}\equiv R^\l_{~\m\l\n}\, , \ee
and with these conventions, the commutator yields
\be
[\nabla_\m,\nabla_\n]V^\a_\b=R^\a_{~\r\m\n}V^\r_\b-R^\r_{~\b\m\n}V^\a_\r\label{c} \, .
\ee
We shall write the integration measure as
\be
d(vol)=\sqrt{|g|} d^n x
\ee
\section{Ward identities from Weyl symmetry.}
Before computing the anomalies for different examples, let us review the origin of the conformal anomaly in the framework that will be used throughout the paper. We are going to use the background field technique combined with the heat kernel approach to compute the counterterms of the differents theories. We expand all the fields of our theory in a background field and a perturbation $\phi_i = \bar{\phi}_i + \phi_i$. When gravity is taken as a fixed background, the quantum fluctuation will be zero, whereas when gravity behaves as a dynamical field, we will have a quantum field associated to the metric $g_{\m\n}= \bg_{\m\n} + \kappa h_{\m\n}$.
\par
Let us examine the symmetries acting on the gravitational field. First of all, there are the diffeomorphisms connected to the identity, which can always be written in a local system of coordinates as
\be
x^\m\rightarrow x^\m+ \xi^\m(x)
\ee
which immediately leads to
\be
\d x^\m= \xi^\m(x)
\ee
There are two interesting ways  of splitting a spacetime diffeomorphism \cite{Weinberg}\cite{Buchbinder}. One of them, called in the jargon
 {\em the background gauge transformations} reads
 \bea
&& \d \bg_{\m\n}=\xi^\l\pd_\l \bg_{\m\n}+\bg_{\l\n}\pd_\m \xi^\l +\bg_{\m\l}\pd_\n\xi^\l =\bn_\m\xi_\n+\bn_\n\xi_\m\nonumber\\
&&\d h_{\m\n}=\xi^\l\pd_\l h_{\m\n}+h_{\l\n}\pd_\m \xi^\l +h_{\m\l} \pd_\n\xi^\l 
\eea
 and a second possibility, dubbed the {\em quantum gauge transformations}  reads
 \bea
 &&\d \bg_{\m\n}=\xi^\l\pd_\l \bg_{\m\n}\nonumber\\
 &&\d h_{\m\n}=\xi^\l\pd_\l h_{\m\n}+\pd_\m \xi^\l \left(\bg_{\l\n}+h_{\l\n}\right)+\pd_\n\xi^\l \left(\bg_{\m\l}+h_{\m\l}\right)
 \eea
 Both classical as well as quantum gauge transformations reproduce the same variation for the spacetime metric, $\d g_{\m\n}$.
 \par
 The second symmetry of interest here is the abelian Weyl symmetry
 \be
 g_{\m\n}(x)\rightarrow \Omega^2(x)\,g_{\m\n}(x)
 \ee
 At the linear level $\Omega=1+\omega(x)$  the metric reads
 \be
 \d g_{\m\n}= 2\omega(x) g_{\m\n}(x)
 \ee
 We represent Weyl's variation of other fields as
 \be
 \d \phi_i=\l_i \phi_i
 \ee
 The advantage of the background field approach is that it is possible to perform the functional integral while preserving background gauge invariance (provided an appropiate gauge fixing is chosen) while breaking the quantum gauge transformations.
 \par
 Working to one loop order, it simplifies 
  \bea
&& \d \bg_{\m\n}=\xi^\l\pd_\l \bg_{\m\n}+\bg_{\l\n}\pd_\m \xi^\l +\bg_{\m\l}\pd_\n\xi^\l =\bn_\m\xi_\n+\bn_\n\xi_\m\nonumber\\
&&\d h_{\m\n}=\xi^\l\pd_\l h_{\m\n}
\eea
 and to
 \bea
 &&\d \bg_{\m\n}=\xi^\l\pd_\l \bg_{\m\n}\nonumber\\
 &&\d h_{\m\n}=\xi^\l\pd_\l h_{\m\n}+\bg_{\l\n}\pd_\m \xi^\l +\bg_{\m\l}\pd_\n\xi^\l 
 \eea
 they still act nonlinearly of the quantum fluctuations owing to the inhomogeneous terms. This physically means that the quantum fluctuations  behave as goldstone bosons of broken diffeomorphism invariance.

 In order to compute the one-loop effective action it is enough to expand the total action up to quadratic order in the quantum fluctuations 
 \be
 S= \overline{S}\left[\bar{\phi}_i, \bg_{\m\n}\right]+\int d(vol) \sum_{ij}\phi_i \Delta_{ij}\left[\bar{\phi}_i, \bg_{\m\n}\right]\phi_j
 \ee
 This leads to the one-loop effective action
\be
\Gamma [\bar{\phi}_i, \bg_{\m\n}] = \overline{S}[\bar{\phi}_i, \bg_{\m\n}] - \dfrac 12 \ln \det \Delta_{ij}[\bar{\phi}_i, \bg_{\m\n}] 
\ee
where we have the determinant of the quadratic operator $\Delta_{ij}[\bar{\phi}_i, \bg_{\m\n}] $. In the heat kernel approach (see \cite{Barvinsky, Vassilevich}) the divergent part of such determinants for minimal operators can be computed via the heat kernel coefficients of the short proper time expansion of the heat kernel
\be
K(\t)\equiv \text{tr}\,e^{-\t \Delta_{ij}}
\ee
\be
\left.K(\t)\right|_{\t\sim 0}=(4\pi\t)^{-{n\over 2}}\,\sum_{n=0}^\infty a_n \t^n
\ee

 In four dimensions we are lead to
\be
\frac{1}{2}\ln \det K=-\frac{1}{4\pi^2}\frac{1}{n-4}\int d^4 x~\sqrt{|g|}~\text{tr}\, a_2 
\ee
Let us introduce an operator implementing the total Weyl variation of any functional
\be 
 {\cal D} \equiv  \sum_i \l_i \bp_i {\d \over \d \bp_i}+2 \bg_{\a\b}{\d\over \d \bg_{\a\b}}
 \ee
 where $\l_i$ corresponds to the scaling dimension of the different fields 
\be
\phi_i \rightarrow \Omega(x)^{\l_i} \, \phi_i.
\ee

The condition for the effective action to be  Weyl invariant then reads
\be
\d \Gamma= \int d(vol) \,  {\cal D}\, \Gamma[\bg_{\m\n},\bp_i]=0
\ee

When there is a violation of this Ward identity there is an anomaly associated to Weyl invariance
\be
{\cal D} \Gamma[\bg_{\m\n},\bp_i]={\cal A} \neq 0
\ee
 It can be shown \cite{EA}, that the anomaly is going to be proportional to the finite part of the one-loop counterterm, that it, to the $a_2$ heat kernel coefficient. In dimensional regularization, the renormalized effective action is defined as
\be
\Gamma_{ren} [\bar{g}_{\mu\nu},\bar{\phi}_i] = \lim_{n \rightarrow 4} \left\lbrace \Gamma[\bar{g}_{\mu\nu},\bar{\phi_i};n] - \Gamma_\infty [\bar{g}_{\mu\nu},\bar{\phi_i}] \right\rbrace
\ee
Here, $\Gamma$ refers to the total one-loop effective action, $\Gamma_{ren}$ to the renormalized effective action and $\Gamma_\infty$ to the one-loop counterterm. 
\par
We have already pointed out the fact that in the background field approach one can fix the quantum gauge transformations while leaving the background ones untouched, so that the one-loop full effective action (which still depends on the background metric) has the same symmetries as the original action
\be
\mathcal{D} \Gamma[\bar{g}_{\mu\nu}, \bar{\phi}_i;n] = \left[2 \bar{g}_{\mu\nu} \dfrac{\delta}{\delta \bar{g}_{\mu\nu}}+\l_i \bar{\phi_i} \dfrac{\delta}{\delta \bar{\phi}}\right] \Gamma[\bar{g}_{\mu\nu}, \bar{\phi}_i;n] = 0
\ee

For the same reason as before, the divergent piece (which is only defined in four dimensions) is Weyl invariant in four dimensions
\be
\mathcal{D} \Gamma_\infty [\bar{g}_{\mu\nu},\bar{\phi}] = (n-4)\,  \Gamma_\infty [\bar{g}_{\mu\nu},\bar{\phi}]
\ee
These leads to the appearance of evanescent operators, the variation of the counterterm being proportional to $(n-4)$,  such that we end up with a finite anomaly which is the residue of the pole of the counterterm
\be
\mathcal{D}\Gamma_{ren}[\bar{g}_{\mu\nu},\bar{\phi}_i]= \mathcal{D} \left[\dfrac{1}{(n-4)} \mathcal{A}[\bar{g}_{\mu\nu},\bar{\phi}_i]\right] = \mathcal{A}[\bar{g}_{\mu\nu},\bar{\phi}]
\ee
Therefore the four dimensional anomaly can be directly computed in terms of the $a_2$ heat kernel coefficient
\be
\delta \Gamma_{ren} =  \int d(vol) \, \mathcal{D}\Gamma_{ren}[\bar{g}_{\mu\nu},\bar{\phi}_i] = \int d(vol) \,\mathcal{A}[\bar{g}_{\mu\nu},\bar{\phi}] =  \frac{1}{4\pi^2} \int d(vol)~\text{tr}\, a_2 \left(x,x\right) 
\ee
\par
 In the next section we will check explicitly  that upon integration by parts, the variation of the effective action equals the heat kernel coefficient. 
\par
With the identification of the anomaly and the heat kernel coefficients \cite{Duff,Bonora,Motola}, it becomes clear that the quantum/classical  character of the gravitational field changes the anomaly, since the counterterm changes due to functional integration upon the gravitational field. 
\par

 Moreover, there is another important difference arising in the dynamical gravity case (that is, when we integrate over ${\cal D} g_{\m\n}$ in the full path integral). This is a consequence of the well known fact \cite{Buchbinder,Kallosh} that in the background field approach counterterms are gauge independent only after the background field equations are substituted in the counterterm (that is,{\em on background shell}). When gravity is functionally integrated over  we have a background  equation of motion  for the gravitational field. 
 \par
 In four dimensions the purely gravitational contribution to the heat kernel coefficients has the form\footnote{The $\Box R$ contribution can be recasted into a counterterm \cite{Duff,Bonora,Motola}.}
\be
\mathcal{A} = c W_4 + a E_4
\ee
where
\bea \bar{W}_4&=&\bR_{\m\n\a\b}^2-2\bR_{\m\n}^2+\frac{1}{3}\bR^2\nonumber\\
\bar{E}_4&=&\bR_{\m\n\a\b}^2-4\bR_{\m\n}^2+\bR^2\eea
correspond to the Weyl squared tensor and the Gauss-Bonnet density (whose integral yields the Euler characteristic, a topological invariant), respectively. 
\par
In the case of a passive gravitational field which is not integrated over, this two pieces appearing in the anomaly are independent.
\par
 On the other hand, when gravity is integrated over, we need to impose the gravitational equations of motion on the counterterm, in order to get a gauge invariant anomaly \cite{Kallosh}. What happens is that this equation of motion relates the two pieces that were independent in the background case. This has direct implications in the classification of anomalies \cite{Deser} and the a-theorem \cite{Komargodski}, which contains information about the coefficient of the Gauss-Bonnet density. 
\section{Scalar field coupled to an abelian  Weyl gauge field.}

As a warm up, let us consider the Weyl invariant action of the scalar field where we have introduced a gauge field associated to Weyl invariance.

\be
S= \int d^4x \, \sqrt{|\bg|}{1\over 2} \, \left[\left(\pd_\m+{\overline W}_\m\right)\phi\right]^2
\ee
 In the spirit of \cite{Gross}, we are interested in comparing the counterterm when the gauge field is just a passive background field without dynamics of its own, and when the gauge field is a full dynamical field. In this case, even if the anomaly changes, there is no coupling of it to the gauge field so that the renormalizability of the theory remains unchanged. 
\par
In order to see the Weyl anomaly of this theory, we need the on-shell divergent part of the effective action which reads
\begin{eqnarray}
\label{count}
\Gamma_\infty[\bg_{\m\n}, \bp,\bW_\m]&=&\frac{1}{n-4}\int d^n x~\sqrt{|\bg|}\,  \frac{1}{(4\pi)^2}\frac{1}{360}\Big[ -\frac{1}{\bp} 60\bar{R}\bar{\Box}\bp + \frac{1}{\bp^2}180 (\bar{\Box}\bp)^2 \nonumber \\
&&+(5\bar{R}^2-2\bar{R}_{\m\n}^2+2\bar{R}_{\m\n\rho\sigma}^2) \Big]\nonumber\\
\end{eqnarray}
In order to eliminate the inverse powers of the scalar field we could use a field redefinition
\be
\phi(x) = e^{D(x)}
\ee
which prevents the scalar field to reach zero value. We shall refrain to do that explicitly here, in order to  not clutter too much the equations that follow.
\par

Let us now analyze the counterterm in a fixed gravitational background but allowing the gauge field to be  a full dynamical field
\be
S= \int d^4x\sqrt{|g|} \left(  -{1\over 4 } F_{\m\n}^2+{1\over 2}\left[\left(\nabla_\m+W_\m\right)\phi\right]^2 \right)
\ee
with $F_{\a\b}=\nabla_\a W_\b-\nabla_\b W_\a$.  
Gravity is again taken as a non dynamical background $\bar{g}_{\m\n}$, but the dynamics of the gauge field is now taken into account.
The equations of motion now read  
\bea
& \Box\bp+\bp\bn_\m\bar{W}^\m -\bar{W}^\m\bar{W}_\m\bp=0\label{eomphi} \nonumber \\
 &\Box\bar{W}_\m-\bn_\l\bn_\m\bar{W}^\l+ \bp\bn_\m\bp+\bar{W}_\m\bp^2=0\label{eomW}
 \eea
To gauge fix the action, we take the  only background Weyl invariant\footnote{Recall that the background field approach, we gauge fix the quantum transformations as we leave the background field transformations untouched. It can be seen that among the gauge conditions of the type $\chi= \bn_\m W^\m+\a \bar{W}_\m W^\m = 0$, the only Weyl invariant one (with respect to the background quantities) is the one corresponding to $\a=-2$.} condition given by
\be
\chi= \bn_\m W^\m-2\bar{W}_\m W^\m = 0
\ee
so that 
\bea
S_{gf} &=& - \dfrac 1 2    \int d(vol) \ \left( \bn_\m W^\m-2\bar{W}_\m W^\m\right)\left( \bn_\n W^\n -2\bar{W}_\n W^\n \right) \nonumber \\
&=& \dfrac{1}{2} \int d^4x \ \sqrt{|\bg|} \ W_\m\left\lbrace   \bn^\n  \bn^\m  -  R^{\m\n} +2\bar{W}^\m\bn^\n-2\bar{W}^\n\bn^\m-2\bn^\m\bar{W}^\n-4\bar{W}^\m\bar{W}^\n\right\rbrace W_\n\nonumber \\
\eea
In this way, we can cancel the non-minimal piece.

After the gauge fixing the quadratic operator reads
\be
S_{2+gf} = \dfrac 1 2 \int \ d(vol) \left[\phi \, M \, \phi + W_\m \, E^\m \, \phi + W_\m \, K^{\m\n} \, W_\n\right] 
\ee
where
\bea
M &=& - \bar{\Box} -  \bn_\m\bar{W}^\m  +  \bar{W}^\m  \bar{W}_\m\nonumber \\
E^\m &=& 2 \bn^\m \bp + 2 \bp \bn^\m +4  \bar{W}^\m \bp \nonumber \\
K^{\m\n} &=&  \bar{\Box} \bg^{\m\n} - R^{\m\n}-(\bn^\m\bar{W}^\n+\bn^\n\bar{W}^\m)-2(\bar{W}^\n\bn^\m-\bar{W}^\m\bn^\n) \nonumber \\
&&-4\bar{W}^\m\bar{W}^\n +\bg^{\m\n} \bp^2 
\eea
Next, we define the generalized field $\Psi^A$ so that we can write
\be
S_2 + S_{gf} =  \int d^4x \ \sqrt{\bar{g}} \ \dfrac 1 2 \Psi^A \left(-C_{AB} \  \bar{ \Box }-{N^\m}_{CB} \  \bar{\nabla}_\m- Y_{AB} \right) \Psi^B
\ee
with
\bea
& \Psi_A = \begin{pmatrix} \phi \\ W_\l \end{pmatrix} \nonumber \\
& \Psi_B = \begin{pmatrix} \phi \\ W_\n \end{pmatrix} \nonumber \\
& C_{AB} = \begin{pmatrix}
	1& 0 \\0 & -  \bg^{\l\n} \end{pmatrix}  \nonumber \\
& {N^\m}_{CB}  = \begin{pmatrix}
	0& \bp\bg^{\m\n}\\-\bp\bg^{\m\l} & 2\bar{W}^\n\bg^{\m\l}-2\bar{W}^\l\bg^{\m\n}\end{pmatrix}  \nonumber \\
& Y_{AB}= \begin{pmatrix}
	-\bar{W}^\m\bar{W}_\m  + \bn_\m\bar{W}^\m & - 2 \bar{W}^\n \bp-\frac{1}{2}\bn^\n \bp   \\ -2 \bar{W}^\l \bp-\frac{1}{2}\bn^\l \bp  & R^{\m\n} +4\bar{W}^\m\bar{W}^\n+(\bn^\m\bar{W}^\n+\bn^\n\bar{W}^\m)- \bp^2 \bg^{\m\n}  \end{pmatrix} 
\eea

We can avoid the terms with one derivative contained in the operator $N^\m$ redefining the quadratic operator as
\be
S_2^{full} = \dfrac 1 2 \int d^4 x  \ \sqrt{|\bg|} \ \Psi^A (-C_{AB} \  \bar{ \Box } - {N^\m}_{AB} \  \bar{\nabla}_\m - Y_{AB}) \Psi^B = \dfrac 1 2 \int d^4x \ \sqrt{|\bg|} \ \Psi_A \ \Delta_{B }^A\ \Psi^B
\ee
where
\be
{\Delta}_{B}^A = - \left(  \left[ \bn_\m \ \d^A_C + \omega^A_{\m C} \right]\left[ \bn^\m \ \d^C_B + \omega^{\m C}_{ B} \right]+ E^A_B\right) = -C^{AC}(C_{CB} \  \bar{ \Box } + {N^\m}_{CB} \  \bar{\nabla}_\m + Y_{CB})
\ee
For that to hold, the new operators are defined as
\bea 
\omega^\m_{AB} &=&\frac{1}{2} \ N^\m_{AB}=\begin{pmatrix} 0& \frac{1}{2}\bp\bg^{\m\n}\\-\frac{1}{2}\bp\bg^{\m\l} & \bar{W}^\n\bg^{\m\l}-\bar{W}^\l\bg^{\m\n}\end{pmatrix} 
\eea
and therefore
\be
S_2 + S_{gf} =  \int d^4x \ \sqrt{\bar{g}} \ \dfrac 1 2 \Psi^A \left(- C_{AB} \tilde{\bar{\Box}} - E_{AB}\right) \Psi^B
\ee
with
\bea
&&E_{AB}= Y_{AB} -\omega_{\m AD} \  \omega^{\m D}_{~~~B}=  \nonumber \\
&&=\begin{pmatrix}
	\bn^\m \bar{W}_\m-\bar{W}^\m\bar{W}_\m -\frac{n}{4} \bp^2   & (n-5)\frac{\bp}{2} \bar{W}^\n  -\frac{1}{2}\bn^\n \bp \\(n-5)\frac{\bp}{2}  \bar{W}^\l \bp -\frac{1}{2}\bn^\l \bp & \bR^{\l\n} +\bn^\l\bar{W}^\n+\bn^\n\bar{W}^\l-(n-6)\bar{W}^\l\bar{W}^\n-\bar{W}_\a\bar{W}^\a\bg^{\l\n}-\frac{3}{4}\bp^2\bg^{\l\n}\end{pmatrix} \nonumber\\
\eea

In order to get the field strength we need to remember that our connection has been modified so that
\be
\left[ D^W_\a, D^W_\b \right] \Psi^A= W^{~~A}_{\a\b~B} \Psi^B
\ee
i.e.
\be W_{\a\b\cdot B}^{~~~~A}=\bar{\nabla}_\a\omega_{\b\cdot B}^{~A}-\bar{\nabla}_\b\omega_{\a\cdot B}^{~A}+\omega_{\a\cdot C}^{~A}\omega_{\b\cdot B}^{~C}-\omega_{\b\cdot C}^{~A}\omega_{\a\cdot B}^{~C}\ee
in our case it is 
\bea
& W^{~~A}_{\a\b~B}= \begin{pmatrix}
	W^{~~\phi}_{\a\b~\phi}&W^{~~\phi}_{\a\b~W}\\ W^{~~W}_{\a\b~\phi} &W^{~~W}_{\a\b~W} \end{pmatrix}\eea
where
\bea\label{W}
W^{~~\phi}_{\a\b~\phi}&&=0\nonumber \\
W^{~~\phi}_{\a\b~W}&&=\frac{1}{2}\left(\d^\n_\b \bn_\a \bp-\d^\n_\a \bn_\b \bp\right)+\frac{\bp}{2}(\d_\b^\n\bar{W}_\a-\d_\a^\n\bar{W}_\b)\nonumber\\
W^{~~W}_{\a\b~\phi}&&=\frac{1}{2}\left(\bg_{\b \l} \bn_\a \bp- \bg_{\a \l} \bn_\b \bp\right)+\frac{\bp}{2}\bp(\bg_{\b \l}\bar{W}_\a-\bg_{\a \l}\bar{W}_\b)\nonumber\\
W^{~~W}_{\a\b~W}&&=\bR^{~\n}_{\l\cdot\a\b}+(\d_\b^\n \bn_\a \bar{W}_\l- \d_\a^\n \bn_\b \bar{W}_\l) +(\bg_{\a \l} \bn_\b \bar{W}^\n- \bg_{\b \l} \bn_\a \bar{W}^\n)+\nonumber\\
&&+\bar{W}^\n(\bg_{\a \l}\bar{W}_\b-\bg_{\b \l}\bar{W}_\a)+\bar{W}_\l(\d_{\b}^\n\bar{W}_\a-\d_{\a}^\n\bar{W}_\b) + (\bg_{\a \l} \d^\n_\b-\bg_{\b \l} \d^\n_\a ) \left(\frac{\bp^2}{4}- W_\m W^\m\right)\nonumber\\
\eea
\par
On the other hand, the ghost action corresponding to the gauge fixing yields
\be
S_{gh}= \dfrac 1 2 \int d^4x \ \sqrt{|\bg|} \ \bar{c} \, \left(-\bar{\Box}+2\bar{W}^\m\bn_\m\right)\,c
\ee
\par
Now, with the usual heat kernel technique, the total counterterm reads
\begin{eqnarray}
	&&\mbox{tr} a_{2}-2\mbox{tr}\, a_{2}^{gh} = \frac{1}{(4\pi)^2}\frac{1}{360}\Big[2(n-17)\bR^2_{\m\n\r\s}-2(n-92)\bR_{\m\n}^2+5(n-14)\bR^2+\nonumber\\
	&&+120(n-4)\bR\bW^\a\bW_\a+120(n-8)(n-4)\bW^\a\bW_\a\bW^\b\bW_\b-120(n-4)\bn_\m\bW^\n\bn^\m\bW_\n+\nonumber\\
	&&480\bn_\m\bW^\n\bn_\n\bW^\m-480\bn_\m\bW^\m\bn_\n\bW^\n+480\bR_{\m\n}\bn^\m\bW^\n-240\bR\bn_\m\bW^\m+\nonumber\\
	&&+(1920-360n)\bR_{\m\n}\bW^\m\bW^\n-960(n-5)\bW^\m\bW^\n\bn_\n\bW_\m+240(n-2)\bW^\m\bW_\m\bn_\n\bW^\n+\nonumber\\
	&&+60\bar{R}(\bn_\m \bar{W}^\m -  \bar{W}^\m  \bar{W}_\m ) + 180(\bn_\m \bar{W}^\m -  \bar{W}^\m  \bar{W}_\m )^2+ (5\bar{R}^2-2\bar{R}_{\m\n}^2+2\bar{R}_{\m\n\rho\sigma}^2) +\nonumber\\
	&& +\dfrac{15}{2} n (n+14)\bp^4+ 30(n-10)\bp^2\bR-30(2n^2 -49n +128)\bp^2 \bW^2+\nonumber\\
	&&+ 240 (n-4)\bp\bar{W}^\a\bn_\a\bp-30 (5n+16)\bp^2\bn_\a\bar{W}^\a+30(n-4)\bn_\m \bp \bn^\m \bp 
	\Big]\nonumber\\
\end{eqnarray}

Finally, the on-shell divergent piece of the effective action reads, in $n=4$ and using the equation of motion of the scalar field \eqref{eomphi} 
\bea
\Gamma_\infty[\bg_{\m\n}, \bp,\bW_\m]&=&\frac{1}{(n-4)}\int d(vol)\frac{1}{(4\pi)^2}\frac{1}{360}\left[\def\bp{\bar{\phi}}-24 \bR^2_{\m\n\r\s} +174 \bR_{\m\n}^2 -45 \bR^2 + 540 \bp^4 \right. \nonumber \\
&&\left.- 60\bar{R}\dfrac{\Box \bp}{\bp} + 180\left(\dfrac{\Box \bp}{\bp}\right)^2 -180 \bR \bp^2 +1080 \bp \Box \bp \right] 
\eea
Comparing the result with the result for a non-dynamical gauge field , it is clear that the anomaly changes, and new terms contribute to it when the gauge field runs in the loops. This is true even in flat space, where it appears the new term
\be
3\frac{1}{(4\pi)^2}\frac{1}{(n-4)}\int d(vol)\, \bigg\{\bp \Box \bp +{1\over 2}\bp^4\bigg\}
\ee
The variation under a local Weyl transformation takes the form 
\bea
\d\Gamma_{\infty} [\bg_{\m\n}, \bp,\bW_\m]&=& \dfrac{1}{n-4} \int d(vol)\, (n-4) \left[-\dfrac{60}{\bp}\left(\omega \bR\bar{\Box}\bp+\bR\bn_\m\omega\bn^\m\bp -2 \bar{\Box}\omega \bar{\Box}\bp \right) \right. \nonumber \\
&+& \left. \dfrac{180}{\bp^2} \left(\omega(\bar{\Box}\bp)^2+2\bn_\m\omega\bn^\m\bp\bar{\Box}\bp\right) + \left(-24 R_{\m\n\r\s}^2+174  R_{\m\n}^2-45  R^2\right) \omega  \right. \nonumber \\
& -&6 \bn_\m \bR  \bn^\m \omega+ \left. 540 \bp^4 -180 \bR \bp^2 + 1080 \bp \bar{\Box} \bp +360 \bp^2 \bar{ \Box }\bp \right] 
\eea
so that it is Weyl invariant in $n=4$ as expected.

\par
\section{Scalar field coupled to Einstein-Hilbert gravity. }

As we have seen in the previous section, the conformal anomaly is sensitive to the character of the fields involved in the theory. After this warm up, we turn on to the main point of this work, namely studying the fate of the Weyl symmetry at the one-loop order  for the different couplings of the fields to gravity, making particular emphasis on the differences between considering it as a classical background or as a quantum dynamical field.
\par

Everything done so far was carried out in a classical non-dynamical gravitational background. We are now interested in studying the effect of quantum dynamical gravity in the Weyl anomaly arising in such theories. 
When gravity is quantum dynamical, even global Weyl invariance imposes severe constraints to correlators. For example, the Green function involving any number of gravitons must vanish, as it does every correlator between fields whose total Weyl charge is nonvanishing. This is exactly the same reason why in ordinary QED all Green functions where the number of $\psi$ fields and the number of $\overline{\psi}$ fields do not match do vanish as well
\be
\left\langle 0_+\left|g_{\m_1\n_1}(x_1)\ldots g_{\m_n \n_n}(x_n)\right|0_-\right\rangle=0
\ee
To be specific, if we denote by $\l_i$ the Weyl charge of the field $\phi_i$; that is,
\be
\d \phi_i =\l_i \phi_i
\ee
The condition that a correlator
\be
\left\langle 0_+\left|\phi^{i_1}(x_1)\ldots \phi^{i_n}(x_n)\right|0_-\right\rangle\neq 0
\ee
does not vanish is that
\be
\sum_{i=1}^{i=n} \l_i=0
\ee
Note in particular that this prevents a cosmological constant to appear upon quantum corrections, since the volume term
\be
\int \sqrt{|g|} d^n x
\ee
is not Weyl invariant (not even globally).
\par
In order to illustrate the effect of the character of the gravitational field in the conformal anomaly, different examples are analyzed. First, we review some of the results of the computations carried out in non-conformal and conformal dilaton gravity \cite{EA}\cite{Nicolai}, which are the only Weyl invariant theories that can be constructed at a linear order in the curvature. 
\par
After that, we compute the anomaly of a scalar field in quadratic gravity theories in order to reinforce these ideas in a different setup. 
\par
The general non-minimal coupling of a scalar field to gravity takes the form
\be S=\int d^n x~\sqrt{|g|}\left[\xi R\phi^2+\frac{1}{2}g^{\m\n}\nabla_\m\phi\nabla_\n\phi\right]\ee
This action is invariant under global Weyl transformations, and moreover, the symmetry is upgraded to local Weyl invariance for the particular value
\be \xi =  \xi_c=\frac{n-2}{8(n-1)}\ee
This enhancement of the gauge group means that the gauge fixing and the ghost sector change accordingly. Next,  we give the results separately
\subsection{Non-Conformal Dilaton Gravity}
Let us start with the simplest case of the non-conformal coupling in the presence of a non dynamical gravitational background
\be S^{\text{\tiny{$\bg$}}}_{\text{\tiny{nCDG}}}=\int d^n x~\sqrt{|\bg|}\left[\xi \bR\phi^2+\frac{1}{2}\bg^{\m\n}\bn_\m\phi\bn_\n\phi\right]\ee
The equation of motion for the scalar field reads
\be \bR=\frac{1}{2\xi}\frac{\bn^2\bp}{\bp}\ee
In order to obtain the counterterm, we compute the determinant of the quadratic operator 
\be
\Delta =- \dfrac 1 2 \bar{ \Box } + \xi \bR 
\ee
This is the first example that proves the dependence on the character of the gravitational field when computing the conformal anomaly. In the non-dynamical case the anomaly reads
\be
\mathcal{A}_{\text{\tiny{nCDG}}}^{\text{\tiny{$\bg$}}}=\frac{1}{(4\pi)^2}\Big[\frac{1}{360}(3\bar{W}_4-\bar{E}_4)+\frac{1}{288}\frac{(1-12\xi)^2}{\xi^2}\frac{(\bn^2\bp)(\bn^2\bp)}{\bp^2} \Big]
\ee
When gravity is a dynamical field, however, one gets that the anomaly changes to
\be
\mathcal{A}_{\text{\tiny{nCDG}}}= \frac{1}{(4\pi)^2}\left[\frac{71}{60}\bar{W}_4+\frac{1259}{1440}\frac{(1-12\xi)^2}{\xi^2}\frac{(\bn^2\bp)(\bn^2\bp)}{\bp^2} \right]
\ee
\subsection{Conformal Dilaton Gravity}
In the critical case $\xi=\xi_c$, the global symmetry is upgraded to the local Weyl symmetry. In the case of non-dynamical gravity, the anomaly can be easily computed, yielding
\be
\mathcal{A}^{\text{\tiny{$\bg$}}}_{\text{\tiny{CDG}}}[\bg_{\m\n}, \bp]=\frac{1}{(4\pi)^2}\Big[\frac{1}{360}(3\bar{W}_4-\bar{E}_4) \Big]
\ee
When gravity is dynamical, there are two local symmetries to be analyzed and gauge fixed in the theory. This imposes further difficulties in the computation, leading to the use of BRST techniques. The detailed computation is carried out in \cite{EA} where they get the result 
\be
\mathcal{A}_{\text{\tiny{CDG}}}[\bg_{\m\n}, \bp]=\frac{1}{(4\pi)^2}\Big[\frac{53}{45}\bar{W}_4 \Big]
\ee
Again, the difference in the anomaly is clearly seen in this example. Moreover, in these examples, the equations of motion  of the gravitational field imposes the equivalence between $\bar{E}_4$ and $\bar{W}_4$, so this is a clear example where the distinction between the two terms vanishes. 

\section{Scalar field coupled to  quadratic gravity.}
In order to reinforce our idea of the modification of the anomaly by dynamical gravitational corrections, we want to take another example of dynamical gravity together with a scalar field. At a linear order in the curvature, the only Weyl invariant possibility is the non minimal coupling $\phi^2 R$ mentioned before, due to the fact already mentioned that in four dimensions the Einstein-Hilbert action is not even invariant under rigid Weyl transformations. 
\par
It seems then that the next easiest alternative is to turn to quadratic theories of gravity, as the quadratic invariants are at least invariant under global conformal transformations. However, in order to illustrate the dependence on the dynamics of the gravitational field in the breaking of the conformal Ward identities, it is enough to consider global Weyl invariant theories.  For convenience in the computation, we take the following quadratic action

\be S_{quad}=\int d^n x~\sqrt{|g|}\left[(\Box \vp)^2+ R^2_{\m\n}-\frac{1}{2}R^2\right]\label{1}\ee
Let us note that we are also taking a quartic derivative term for the scalar field\footnote{At a technical level, this choice is made in order to have the same number of derivatives acting on both fields. This leads to a determinant of a minimal operator, which simplifies the computations.}. In this case, the scalar field is inert under global Weyl transformations, as opposed to the Weyl assignment of the previous example, and there is no need of introducing the gauge field associated to the gauging of the symmetry. 
\par
We first want to compute the conformal anomaly in the presence of a gravitational background field. The equation of motion for the scalar field reads
\bea
&&{\d S\over \d \vp}=\bar{\Box}^2 \bvp=0\eea
Up to quadratic order in the expansion of the scalar field, we need to compute the determinant of the operator
\be
S_2= \kappa^2 \int d^4x \,  \sqrt{|\bg|} \vp \bar{\Box}^2 \vp
\ee
Let us stress that when gravity is 
classical, the scalar field expansion is unaffected by the pure gravitational part, so that it does not appear any contribution in the quadratic operator. 
In this simple case we just have to compute the determinant of the $\bar{\Box}^2$ operator. This is tabulated in \cite{Barvinsky} and finally we obtain the anomaly
\bea
\mathcal{A}^{\bg}_{\text{\tiny{quad}}}[\bg_{\m\n}, \bvp]&=&\frac{1}{(4\pi)^2}\frac{1}{180}\Big[2\bR_{\m\n\a\b}^2-2\bR_{\m\n}^2+5\bR^2 \Big]
\label{quadnondynam}
\eea

\par
Turn now to the case where the gravitational field is quantum dynamical, so that the background field expansion reads\bea
g_{\m\n}&=&\bg_{\m\n}+ \kappa h_{\m\n}\nonumber\\
\vp&=&\bvp+ \kappa \vp
\eea
The equations of motion for the graviton obtained by demanding that the  linear order of the expansion vanishes.
\bea 
&&-\frac{1}{2}\bg_{\m\n} \bR_{\a\b}\bR^{\a\b}+2\bR_{\m\l}\bR_\n^{\,\l}+\frac{1}{4}\bg_{\m\n} \bR^2-\bR_{\m\n}\bR-{1\over 2} \bg_{\m\n}\bar{ \Box }\bvp \bar{ \Box }\bvp +\bn_{\mu}\bvp \bn_\n \bar{ \Box }\bvp \nonumber \\
&&+\bn_{\nu}\bvp \bn_\m \bar{ \Box }\bvp-\bg_{\m\n} \bn^\l \bvp \bn_\l \bar{ \Box } \bvp =0\eea

Let us now restrict ourselves to constant curvature backgrounds from now on since it is enough   to illustrate the effect of quantum dynamical gravity in the anomaly while making the computations  more tractable. The main reason for this stems from the  fact that after integrating by parts, several terms in the action simplify due to the constancy of the background curvature. With our conventions, for constant curvature spaces we have 
\be \bR_{\a\b\m\n}=-\frac{\l}{3}\left(\bg_{\a\m}\bg_{\b \n}-\bg_{\a\n}\bg_{\b \m}\right)\ee
\par
In order to compute the one-loop counterterm which will give us the conformal anomaly, we need to compute the determinant of the quadratic operator. In this case, up to quadratic order in the fluctuations we have
\be
S_2 = \int \ d^n x \ \sqrt{\bar{g}} \left[h^{\a\b}M_{\a\b\g\e}h^{\g\e}+h^{\a\b} N_{\a\b}\vp+\vp K\vp\right] 
\ee
where the different operators are 
\bea\label{K}
M_{\a\b\g\e}&=& \kappa^2 \Big[\frac{1}{4}\bg_{\a\b}\bar{\Box}\bn_\g\bn_\e+\frac{1}{4}\bg_{\g\e}\bn_\a\bn_\b\bar{\Box}-\frac{1}{4}\bg_{\a\g}\bn_\b\bn_\e\bar{\Box}-\frac{1}{4}\bg_{\a\g}\bn_\e\bn_\b\bar{\Box}+\frac{1}{4}\d_{\a\g,\b\e}\bar{\Box}^2\nonumber \\
&&
-\frac{1}{4}\bg_{\a\b} \bg_{\g\e}\bar{\Box}^2-\l\big[-\frac{1}{2}\d_{\a\g,\b\e}\bar{\Box}+\frac{1}{6}\left(\bg_{\a\g}\bn_\b\bn_\e+\bg_{\a\e}\bn_\b\bn_\g+\bg_{\b\g}\bn_\a\bn_\e+\bg_{\b\e}\bn_\a\bn_\g\right)+\nonumber\\
&&+\frac{1}{6}\left(\bg_{\a\g}\bn_\e\bn_\b+\bg_{\a\e}\bn_\g\bn_\b+\bg_{\b\g}\bn_\e\bn_\a+\bg_{\b\e}\bn_\g\bn_\a\right) - \dfrac{1}{6} \left(\bg_{\m\n}\bn_\g\bn_\e+\bg_{\g\e}\bn_\a\bn_\b\right)\big] \nonumber \\
&&+\l^2\big[-\frac{4}{9}\d_{\a\g,\b\e}+\frac{1}{9}\bg_{\a\b}\bg_{\g\e}\big] + \nonumber \\
&&+\frac{1}{8}\left(\bg_{\a\b}\bg_{\g\e}-\bg_{\a\g}\bg_{\b\e}-\bg_{\a\e}\bg_{\b\g}\right)(\bar{ \Box }\bvp)^2+\frac{1}{4}\left(\bg_{\a\g}\bg_{\b\e}+\bg_{\a\e}\bg_{\b\g}\right)\left[(\bar{ \Box }\bvp)^2+\bn_\l\bvp\bn^\l\bar{ \Box }\bvp\right]+\nonumber\\
&&+\frac{1}{4}\bg_{\a\b}\bg_{\g\e}\left(\bn_\l\bvp\bar{ \Box }\bvp\bn^\l-\bn_\t\bvp\bn^\t\bn_\l\bvp\bn^\l-\bn^\t\bvp\bn_\l\bvp\bn_\t\bn^\l\right)+\nonumber\\
&&+\frac{1}{4}\bg_{\a\b}\left(\bn_\g\bvp\bn_\e\bar{ \Box }\bvp+\bn_\g\bvp\bn_\l\bvp\bn_\e\bn^\l+\bn_\g\bvp\bn_\e\bn_\l\bvp\bn^\l\right)+\nonumber\\
&&+\frac{1}{4}\bg_{\g\e}\left(\bn_\a\bvp\bn_\b\bar{ \Box }\bvp+\bn_\a\bvp\bn_\l\bvp\bn_\b\bn^\l+\bn_\a\bvp\bn_\b\bn_\l\bvp\bn^\l\right)-\nonumber\\
&&+\frac{1}{4}\bg_{\a\b}\left(\bn_\e\bvp\bn_\g\bar{ \Box }\bvp+\bn_\e\bvp\bn_\l\bvp\bn_\g\bn^\l+\bn_\e\bvp\bn_\g\bn_\l\bvp\bn^\l\right)+\nonumber\\
&&+\frac{1}{4}\bg_{\g\e}\left(\bn_\b\bvp\bn_\a\bar{ \Box }\bvp+\bn_\b\bvp\bn_\l\bvp\bn_\a\bn^\l+\bn_\b\bvp\bn_\a\bn_\l\bvp\bn^\l\right)-\nonumber\\
&&-\frac{1}{2}\left(\bg_{\a\g}\bn_\b \bvp \bn_\e\bar{ \Box }\bvp+\bg_{\a\e}\bn_\b \bvp \bn_\g\bar{ \Box }\bvp+\bg_{\b\g}\bn_\a \bvp \bn_\e\bar{ \Box }\bvp+\bg_{\b\e}\bn_\a \bvp \bn_\g\bar{ \Box }\bvp\right)+\nonumber\\
&&+\bn_\a\bn_\b\bvp\bn_\g\bn_\e\bvp+\dfrac 1 4 (\bn_\b\bvp\bn_\g\bn_\e\bvp\bn_\a+\bn_\a\bvp\bn_\g\bn_\e\bvp\bn_\b+\bn_\e\bvp\bn_\a\bn_\b\bvp\bn_\g \nonumber \\
&&+\bn_\g\bvp\bn_\a\bn_\b\bvp\bn_\e)- \dfrac 1 8 (\bn_\b\bvp\bn_\e\bvp\bn_\a\bn_\g+\bn_\a\bvp\bn_\e\bvp\bn_\b\bn_\g+\bn_\b\bvp\bn_\g\bvp\bn_\a\bn_\e+ \nonumber \\
&&+\bn_\a\bvp\bn_\g\bvp\bn_\b\bn_\e + \bn_\b\bvp\bn_\e\bvp\bn_\g\bn_\a+\bn_\a\bvp\bn_\e\bvp\bn_\g\bn_\b+\bn_\b\bvp\bn_\g\bvp\bn_\e\bn_\a \nonumber \\
&&+\bn_\a\bvp\bn_\g\bvp\bn_\e\bn_\b)-\dfrac 1 8 (\bn_\g\bvp\bn_\b\bn_\e\bvp\bn_\a+\bn_\g\bvp\bn_\a\bn_\e\bvp\bn_\b+\bn_\a\bvp\bn_\b\bn_\e\bvp\bn_\g+ \nonumber \\
&&+\bn_\e\bvp\bn_\a\bn_\g\bvp\bn_\b + \bn_\e\bvp\bn_\b\bn_\g\bvp\bn_\a+\bn_\e\bvp\bn_\a\bn_\g\bvp\bn_\b+\bn_\a\bvp\bn_\b\bn_\g\bvp\bn_\e+\nonumber \\
&& +\bn_\g\bvp\bn_\a\bn_\e\bvp\bn_\b) - \dfrac{\l}{12} (2 (\bg_{\a \g} \bn_\b \bvp \bn_\e \bvp+\bg_{\b \g} \bn_\a \bvp \bn_\e \bvp+\bg_{\a \e} \bn_\b \bvp \bn_\g \bvp+\bg_{\a \g} \bn_\b \bvp \bn_\e \bvp )  \nonumber \\
&&- (\bg_{\g\e} \bn_\a \bvp\bn_\b \bvp+ \bg_{\a\b} \bn_\g \bvp\bn_\e \bvp))
\Big] \nonumber\\
N_{\a\b} &=& \kappa^2\Big[-\bg_{\a\b}\left(\bar{ \Box }\bvp\bar{ \Box }+\bn_\l\bar{ \Box }\bvp\bn^\l+\bn^\l\bvp\bn_\l\bar{ \Box }\right)+\bn_\a\bar{ \Box }\bvp\bn_\b+ \bn_\b\bar{ \Box }\bvp\bn_\a +\bn_\b\bvp\bn_\a\bar{ \Box } + \nonumber \\
&&+ \bn_\b\bar{ \Box }\bvp\bn_\a\Big]\nonumber\\
K& =&\kappa^2  \bar{\Box}^2   \nonumber \\  
\eea
\par
The only local symmetry that need to be gauge fixed is diffeomorphism invariance. In order to do so, we  choose some gauge fixing condition $\chi_\m$ and a weight operator $G_{\m\n}$ \cite{Buchbinder} (in order to account for the terms with four derivatives), so that 
\bea S_{gf}&=& \kappa^2\int d^4x \ \sqrt{|g|}\chi_\m G^{\m\n}\chi_\n
\eea
where
\bea
\chi_\m&=&\nabla_{\r}h^\r_\m-\frac{1}{2}\nabla_\m h-2\vp\bn_\m \bvp\nonumber\\
G^{\m\n}&=&\frac{1}{2}\left(-g^{\m\n}\Box-\nabla^\m\nabla^\n+\nabla^\n\nabla^\m\right)
\eea
In constant curvature spacetimes we have 
\be
S_{gf} =\int \ d^n x \ \sqrt{\bar{g}}\left[h^{\a\b}M^{gf}_{\a\b\g\e}h^{\g\e}+h^{\a\b} N^{gf}_{\a\b}\vp +\vp K^{gf} \vp\right] 
\ee
and the gauge fixing operators take the form
\bea M^{gf}_{\a\b\g\e}&=&\kappa^2 \Big[-\frac{1}{4}\bg_{\a\b}\bar{\Box}\bn_\g\bn_\e-\frac{1}{4}\bg_{\g\e}\bn_\a\bn_\b\bar{\Box}+\frac{1}{4}\bg_{\a\g}\bn_\b\bn_\e\bar{\Box}+\frac{1}{4}\bg_{\a\g}\bn_\e\bn_\b\bar{\Box}+\frac{1}{8}\bg_{\a\b}\bg_{\g\e}\bar{\Box}^2+\nonumber\\
&&-\l\big[\frac{1}{12}\bg_{\a\b}\bg_{\g\e}\bar{\Box}-\frac{1}{3}\d_{\a\g,\b\e}\bar{\Box}-\frac{1}{6}\left(\bg_{\a\g}\bn_\b\bn_\e+\bg_{\a\e}\bn_\b\bn_\g+\bg_{\b\g}\bn_\a\bn_\e+\bg_{\b\e}\bn_\a\bn_\g\right)-\nonumber\\
&&-\frac{1}{6}\left(\bg_{\a\g}\bn_\e\bn_\b+\bg_{\a\e}\bn_\g\bn_\b+\bg_{\b\g}\bn_\e\bn_\a+\bg_{\b\e}\bn_\g\bn_\a\right)+\frac{1}{6}\left(\bg_{\m\n}\bn_\g\bn_\e+\bg_{\g\e}\bn_\a\bn_\b\right)\big]+\nonumber\\
&&+\l^2\big[\frac{8}{9}\d_{\a\g,\b\e}-\frac{2}{9}\bg_{\a\b}\bg_{\g\e}\big]\Big]\nonumber\\
N^{gf}_{\a\b}&=&\kappa^2\Bigg[\bg_{\a\b}\left(\bar{ \Box }\bvp\bar{ \Box }+\bar{ \Box }\bn_\l\bvp\bn^\l+\bn^\l\bvp\bn_\l\bar{ \Box }+\bar{ \Box }^2\bvp+2\bn_\l\bn_\t\bvp\bn^\t\bn^\l+2\bn_\l\bar{ \Box }\bvp\bn^\l\right)-\nonumber\\
&&-\left(\bar{ \Box }\bn_\a\bvp\bn_\b+\bn_\b\bvp\bn_\a\bar{ \Box }+\bn_\a\bn_\b\bar{ \Box }\bvp+\bn_\a\bn_\b\bvp\bar{ \Box }+2\bn_\l\bn_\a\bvp\bn^\b\bn_\l+2\bn_\a\bn_\b\bn_\l\bvp\bn^\l\right) \nonumber \\
&&-\left(\bar{ \Box }\bn_\b\bvp\bn_\a+\bn_\a\bvp\bn_\b\bar{ \Box }+\bn_\b\bn_\a\bar{ \Box }\bvp+\bn_\b\bn_\a\bvp\bar{ \Box }+2\bn_\l\bn_\b\bvp\bn^\a\bn_\l+2\bn_\b\bn_\a\bn_\l\bvp\bn^\l\right)\Bigg]\nonumber\\
K^{gf}&=&\kappa^2\left[-2\bn_\l\bvp\left(\bn^\l\bvp\bar{ \Box }+2\bn^\t\bn^\l\bvp\bn_\t+\bar{ \Box }\bn^\l\bvp\right)\right]\eea
This gauge fixing is enough to obtain a minimal operator. The divergent part of the determinant can be obtained using the heat kernel method. After gauge fixing, the quadratic action takes the form 
\be
S_2+S_{gf} =\int d^4x \ \sqrt{\bar{g}} \  \Psi^A \left(C_{AB} \  \bar{ \Box }^2+{D^{\m\n}}_{AB} \  \bar{\nabla}_\m\bn_\n+{H^\m}_{AB} \  \bar{\nabla}_\m+Y_{AB} \right) \Psi^B
\ee
where we define the generalized field as 
with
\bea
& \Psi_A = \begin{pmatrix} h^{\a\b}\\
	\vp  \end{pmatrix} \nonumber \\
& \Psi_B = \begin{pmatrix} h^{\g\e}\\ \vp\end{pmatrix} \eea

The components of the rest of the operators read
\bea
& C_{AB} =  \kappa^2\begin{pmatrix}
	\dfrac{1}{8}\left(-\bg_{\a\b} \bg_{\g\e} +\bg_{\a\g}\bg_{\b\e}+\bg_{\a\e}\bg_{\b\g}
	\right)&0 \\0 &1 \end{pmatrix}  \nonumber \\
& C^{AB} =  \dfrac{1}{\kappa^2}\begin{pmatrix}
	-\dfrac{4}{(n-2)}\bg_{\a\b} \bg_{\g\e} + 2\left(\bg_{\a\g}\bg_{\b\e}+\bg_{\a\e}\bg_{\b\g}
	\right)&0 \\0 &1 \end{pmatrix}
\eea
and
\bea
D^{\m\n}_{hh}&=&\kappa^2\Bigg[ -\dfrac{\l}{4}\left(\dfrac{1}{3}\bg_{\a\b}\bg_{\g\e}-\dfrac{5}{3}\left(\bg_{\a\g}\bg_{\b\e}+\bg_{\a\e}\bg_{\b\g}
\right) \right) \bg^{\m\n}-\dfrac 1 4\bg_{\a\b}\bg_{\g\e}\bn^\n\bvp\bn^\m\bvp\nonumber\\
&+&\frac{1}{8}\bg_{\a\b}\left(\d^\n_\e\bn^\m\bvp\bn_\g\bvp+\d^\n_\g\bn^\m\bvp\bn_\e\bvp+\d^\m_\e\bn^\n\bvp\bn_\g\bvp+\d^\m_\g\bn^\n\bvp\bn_\e\bvp\right)+\nonumber\\
&+&\frac{1}{8}\bg_{\g\e}\left(\d^\n_\a\bn^\m\bvp\bn_\b\bvp+\d^\n_\b\bn^\m\bvp\bn_\a\bvp+\d^\m_\a\bn^\n\bvp\bn_\b\bvp+\d^\m_\b\bn^\n\bvp\bn_\a\bvp\right)-\nonumber\\
&-& \dfrac 1 8 \left(\d_\a^\m\d_\g^\n\bn_\b\bvp\bn_\e\bvp+\d_\b^\m\d_\g^\n\bn_\a\bvp\bn_\e\bvp+\d_\a^\m\d_\e^\n\bn_\b\bvp\bn_\g\bvp+\d_\b^\m\d_\e^\n\bn_\a\bvp\bn_\g\bvp \right. \nonumber \\
&& \left. +\d_\g^\m\d_\a^\n\bn_\b\bvp\bn_\e\bvp+\d_\g^\m\d_\b^\n\bn_\a\bvp\bn_\e\bvp+\d_\e^\m\d_\a^\n\bn_\b\bvp\bn_\g\bvp+\d_\e^\m\d_\b^\n\bn_\a\bvp\bn_\g\bvp\right)\Bigg]\nonumber\\
D^{\m\n}_{h\vp}&=&\kappa^2\left[\bg_{\a\b}\bn^\m\bn^\n\bvp-\bg^{\m\n}\bn_\a\bn_\b\bvp-\frac{1}{2}\left(\d_\a^\m\bn_\b\bn^\n\bvp+\d_\b^\m\bn_\a\bn^\n\bvp+\d_\a^\n\bn_\b\bn^\m\bvp+\d_\b^\n\bn_\a\bn^\m\bvp\right)\right]\nonumber\\
D^{\m\n}_{\vp h}&=&\kappa^2\left[\bg_{\g\e}\bn^\m\bn^\n\bvp-\bg^{\m\n}\bn_\g\bn_\e\bvp-\frac{1}{2}\left(\d^\n_\e\bn^\m\bn_\g\bvp+\d^\n_\g\bn^\m\bn_\e\bvp+\d^\m_\e\bn^\n\bn_\g\bvp+\d^\m_\g\bn^\n\bn_\e\bvp\right)\right]\nonumber\\
D^{\m\n}_{\vp \vp}&=&-2\kappa^2\bg^{\m\n}\bn_\l\bvp\bn^\l\bvp
\eea
For ${H^\m}_{AB}$ we have 
\bea
H^{\m}_{hh}&=&H^{\m}_{\vp \vp}=0\nonumber\\
H^{\m}_{h\vp}&=&- H^{\m}_{\vp h}= \frac{\kappa^2}{2}\left[\d_\b^\m\bar{ \Box }\bn_\a\bvp+\d_\a^\m\bar{ \Box }\bn_\b\bvp - \l (\bg_{\a\b}  \bn^\m \bvp-\d_\b^\m\bn_\a\bvp-\d_\a^\m\bn_\b\bvp)   + 2 \bg_{\a\b} \bn^\mu \bar{ \Box } \bvp \right. \nonumber \\
&& \left.-2 \bg_{\a\b} \bar{ \Box } \bn^\mu \bvp-2  \bn_\a \bn_\b \bn^\m \bvp+2 \bn^\m \bn_\a \bn_\b  \bvp  \right]\nonumber\\
\eea 
And finally, the operator containing the terms without derivatives takes the form
\bea
Y_{hh}&=&\kappa^2\Bigg[\frac{1}{8}\left(\bg_{\a\g}\bg_{\b\e}+\bg_{\a\e}\bg_{\b\g}\right)(\bar{ \Box }\bvp)^2+\frac{1}{4}\left(\bg_{\a\g}\bg_{\b\e}+\bg_{\a\e}\bg_{\b\g}\right)\bn_\l\bvp\bn^\l\bar{ \Box }\bvp+\frac{1}{8}\bg_{\a\b}\bg_{\g\e} (\bar{ \Box }\bn_{\mu}\bvp \bn^\mu \bvp\nonumber\\
&&-\bn_{\mu}\bar{ \Box }\bvp \bn^\mu \bvp)+\frac{1}{8}\bg_{\a\b}\bg_{\g\e}\bn_\l\bn_\t\bvp\bn^\l\bn^\t\bvp+\frac{1}{4}\bg_{\a\b}\bn_\g\bvp\bn_\e\bar{ \Box }\bvp+\frac{1}{4}\bg_{\g\e}\bn_\a\bvp\bn_\b\bar{ \Box }\bvp-\nonumber\\
&&-\frac{1}{2}\left(\bg_{\a\g}\bn_\b \bvp \bn_\e\bar{ \Box }\bvp+\bg_{\a\e}\bn_\b \bvp \bn_\g\bar{ \Box }\bvp+\bg_{\b\g}\bn_\a \bvp \bn_\e\bar{ \Box }\bvp+\bg_{\b\e}\bn_\a \bvp \bn_\g\bar{ \Box }\bvp\right)+\nonumber\\
&& - \dfrac 1 4 (\bg_{\a\b} \bn_\g \bn_{\mu} \bvp \bn_\e \bn^\mu \bvp +\bg_{\g\e} \bn_\a \bn_{\mu} \bvp \bn_\b \bn^\mu \bvp) - \dfrac 1 8 (\bg_{\a\b} \bn_\g  \bvp \bar{ \Box }\bn_\e  \bvp +\bg_{\g\e} \bn_\a  \bvp \bar{ \Box } \bn_\b  \bvp) \nonumber \\
&&+ \dfrac{1}{4} (\bn_\a\bn_\g\bvp\bn_\b\bn_\e\bvp+\bn_\a\bn_\e\bvp\bn_\b\bn_\g\bvp)+ \dfrac 1 2 \bn_\a\bn_\b\bvp\bn_\g\bn_\e\bvp+\l^2\left(\frac{4}{9}\d_{\a\g,\b\e}-\frac{1}{9}\bg_{\a\b}\bg_{\g\e}\right) \nonumber \\
&& - \dfrac{\l}{12} (2 (\bg_{\a \g} \bn_\b \bvp \bn_\e \bvp+\bg_{\b \g} \bn_\a \bvp \bn_\e \bvp+\bg_{\a \e} \bn_\b \bvp \bn_\g \bvp+\bg_{\a \g} \bn_\b \bvp \bn_\e \bvp ) - (\bg_{\g\e} \bn_\a \bvp\bn_\b \bvp \nonumber \\
&&+ \bg_{\a\b} \bn_\g \bvp\bn_\e \bvp)) +\frac{1}{4}\bg_{\a\b}\bn_\e\bvp\bn_\g\bar{ \Box }\bvp+\frac{1}{4}\bg_{\g\e}\bn_\b\bvp\bn_\a\bar{ \Box }\bvp - \dfrac 1 8 (\bg_{\a\b} \bn_\e  \bvp \bar{ \Box }\bn_\g  \bvp +\bg_{\g\e} \bn_\b  \bvp \bar{ \Box } \bn_\e  \bvp)
\Bigg]
\nonumber\\
Y_{h\vp}&=&= Y_{\vp h}= - \dfrac{\l \, \kappa^2 }{12} (\bar{ \Box } \bvp +14 \bn_\a \bn_\b \bvp )\nonumber\\
Y_{\vp \vp}&=&2\kappa^2\bn_\l\bn_\t\bvp\bn^\t\bn^\l\bvp
\eea

We need to compute the expansion coefficient giving the infinite part of the one-loop effective \cite{Barvinsky} action
\bea
\text{tr}\, a_2 \left(x,x\right)&=&  \frac{1}{(4\pi)^2}\text{tr}\left\{\frac{1}{180}\left(2\bR_{\m\n\r\s}^2-2\bR_{\m\n}^2+5\bR^2\right) \mathbbm{1}+\right.\nonumber\\
&&\left.+\frac{1}{6}W_{\m\n}W^{\m\n}-Y-\frac{1}{6}D^{\m\n}\bR_{\m\n}+\frac{1}{12}D\bR+\frac{1}{48}D^2+\frac{1}{24}D^{\m\n}D_{\m\n}\right\}\nonumber\\
\eea
where the field strength is defined through
\be
[\bn_\m,\bn_\n]h^{\a\b}=W_{\r\s\m\n}^{\a\b}h^{\r\s}
\ee
Then in this case, we find
\bea
\text{tr}\, a_2&=&\frac{1}{(4\pi)^2}\Bigg[\l^2\frac{n}{1620(n-2)^2}\left(-13n^5+793n^4-3653n^3+3210n^2+8100n-8248\right)+\nonumber\\
&+&\frac{n(n+2)}{6}\bn_\l\bvp\bn^\l\bvp\bn_\t\bvp\bn^\t\bvp+\l\frac{n^3 -4n^2+20n-56}{18(n-2)}\bn_\l\bvp\bn^\l\bvp-\nonumber\\
&-&\frac{n^3-5n^2+n+8}{n-2}\bn_\l\bar{ \Box }\bvp\bn^\l\bvp+\frac{(n-4)}{n-2}\bar{ \Box }\bn_\l\bvp\bn^\l\bvp\nonumber\\
&+&\frac{-3n^3+2n^2+4n+24}{6(n-2)}(\bar{ \Box }\bvp)^2+\frac{2n^3+12n^2-56n+48}{12(n-2)}\bn_\l\bn_\t\bvp\bn^\l\bn^\t\bvp\Bigg]
\eea
where we have particularized for constant curvature spaces.
\par
There is another piece of the computation coming from the ghost action after the gauge fixing of diffeomorphism invariance. The ghost action for the previous gauge fixing is 
\bea
S_{gh}^D&=&\int d^4x \ \sqrt{|g|}\bar{C}_\a M^\a_{\b}C^\b\eea
where 
\be M^\a_{\b}=G^{\a\l}\left[\frac{\d\chi_\l}{\d h_{\m\n}}\left(\bg_{\m\b}\bn_\n+\bg_{\n\b}\bn_\m\right)+\frac{\d\chi_\l}{\d C}\bn_\b\bvp\right]\ee
Given that,
\bea \frac{\d\chi^\a}{\d h_{\m\n}}&=&\frac{1}{2}\left(\bg^{\m\a}\bn^\n+\bg^{\n\a}\bn^\m-\bg^{\m\n}\bn^\a\right)\nonumber\\
\frac{\d\chi_\l}{\d C}&=&-2\bn_\l\bvp\eea
and 
\be M_{\a\b}=-g_{\a\b} \bar{ \Box }^2 + R_{\l\a} R_\b^\l-\bR_{\t\b\a\l}\bn^\l\bvp\bn^\t\bvp+\bar{ \Box }(\bn_\a\bvp\bn_\b\bvp)\ee
the ghost action takes the form
\bea
S_{gh}^D&=&\int d^4x \ \sqrt{|g|}\bar{C}^\a \left[-g_{\a\b} \bar{ \Box }^2 + R_{\l\a} R_\b^\l-\bR_{\t\b\a\l}\bn^\l\bvp\bn^\t\bvp+\bar{ \Box }(\bn_\a\bvp\bn_\b\bvp)\right]C^\b \nonumber \\ \eea
the heat kernel coefficient coming from the ghost action is
\bea
\text{tr}\, a^{\text{ghD}}_2 \left(x,x\right)&=&\frac{1}{(4\pi)^2}\frac{n}{1620}\left(49n^2-82n+1680\right) \l^2-\frac{\l}{180}\bn_{\mu}\bvp\bn^\m\bvp
\eea
In constant curvature spaces and for $n=4$ the counterterm reads
\bea
\Gamma_{\infty}[\bg_{\m\n}, \bvp]&=&\frac{1}{(n-4)}\int d^n x~\sqrt{|\bg|}\frac{1}{(4\pi)^2}\Bigg[\dfrac{398\l^2}{45}+\dfrac{61\l}{90}\bn_\m\bvp\bn^\m\bvp+4\bn_\m\bvp\bn^\m\bvp\bn_\n\bvp\bn^\n\bvp+\nonumber\\
&&+2\bn_\m\bar{ \Box }\bvp\bn^\m\bvp+\dfrac{1}{45}\bar{ \Box }\bn_\m\bvp\bn^\m\bvp-10(\bar{ \Box }\bvp)^2+\dfrac{271}{45}\bn_\m\bn_\n\bvp\bn^\m\bn^\n\bvp\Bigg]\nonumber\\
\eea
the on-shell value  is
\bea
\Gamma_{\infty}[\bg_{\m\n}, \bvp]=\frac{1}{(n-4)}\int d^n x~\sqrt{|\bg|}\frac{1}{(4\pi)^2}\left[\dfrac{398\l^2}{45}+\dfrac{61\l}{90}\bn_\m\bvp\bn^\m\bvp+4\bn_\m\bvp\bn^\m\bvp\bn_\n\bvp\bn^\n\bvp\right]
\eea
We can clearly see the difference with the result obtained for non-dynamical gravity \eqref{quadnondynam}, which in constant curvature spacetimes reads
\bea
\mathcal{A}_{\text{\tiny{quad}}}[\bg_{\m\n}, \bvp]&=&\frac{1}{(4\pi)^2}\frac{58}{135}\l^2
\eea
 In the dynamical case, we have new terms appearing which depend on the scalar field. The equation of motion for the graviton does not help in simplifying those terms.  This is the main original result of this work.
\section{Conclusions: quantum  versus classical gravity}
The main aim of the computation of the conformal anomaly in the different examples that we have studied was precisely to point out the differences that arise due to the classical background or dynamical character of gravity. 
\par

The key point is that not only the numerical factors change depending on the dynamical or non-dynamical character of the gravitational field, but also the sign of the anomaly. It is remarkable, for instance, that when dynamical gravity is considered in the non-minimal coupling of the scalar field to gravity, the anomaly vanishes for constant curvature spacetimes, whereas it yields a non-zero contribution when it is considered as a non-dynamical background.

\par

We also point out that in the quantum  gravitational case, the distinction between the Weyl squared term and the Gauss-Bonnet density appearing in the  Weyl anomaly does not hold because of the gravitational equation of motion, which is needed in the counterterm in order to enforce gauge invariance. That is, the difference in the anomaly is not only due to the apparition of gravitons in the loop.
\par
Our main original contribution is the  consideration of quadratic theories of gravity in order to examine the physical consequences of a quantum dynamical gravitational  field in some specific models with global Weyl invariance. The main reason for considering quadratic theories  is that, even if one relaxes the condition of local Weyl invariance to a global one, the only linear (in the Riemann tensor) theory is constructed out of the non-minimal $\phi^2 R$ coupling of the dilaton.  Let us remark that in spite of the unitarity problems that afflict these theories in the second order formalism, it is reasonable to expect that the general form of the Ward identities is not affected by them.
\par
A quite significant implication of our analysis is that it suggests the possibility that by experimentally observing consequences of the anomaly, evidence could be gathered on the quantum/classical dynamical character of the gravitational field. We are only aware of two cases where the difference in the conformal anomaly triggered by the dynamical character of gravity could play a role.  Unfortunately both seem quite far from direct experimental verification in a real quantum gravity setup.
\par
For example  our results modify the  trace anomaly driven, Starobinsky type of inflationary models \cite{Starobinsky}, as they are precisely based on the non-minimal coupling of a scalar field to gravity.
\par
 Moreover, it is well know that chiral anomalies contain gravitational corrections giving rise to mixed axial-gravitational anomalies \cite{Delbourgo}. These have been experimentally {\em observed} in transport effects involving Weyl semimetals \cite{Gooth}, where the low energy quasiparticles behave as fermions. Some obvious caveats are in order here. The condensed matter models that lead to this identification have a well specified range of validity (namely, the existence of quasiparticles). This is not in any sense a substitute to real quantum gravity setups.
   Recent work regarding similar transport effects due to conformal anomalies \cite{Chernodub,Arjona} points towards the possibility of  experimental access to the trace anomaly, which is also expected to be modified by quantum dynamical gravitational effects.
 \par
The contribution of the conformal anomaly to the mass of the hydrogen atom \cite{Sun} has also been computed. Unfortunately, the precision of the experimental results is not enough to check this effect for the time being.

\par 
Unfortunately, we cannot conclude anything qualitatively new concerning the search of a consistent local Weyl invariant theory including the gravitational field. Our analysis only supports the idea that if there were to exist such a quantum field theory, it would be a wondrous thing indeed. This being ultimately related with the difficulty of fulfilling the constraints from diffeomorphism invariance and those from local Weyl invariance \cite{Conformalweyl}.  In fact, such a theory must be a finite one, with vanishing cosmological constant.
\par
 All we know is that there is a family of theories, the self-mirror ones in the Duff-Ferrara sense \cite {DuffFerrara} for which the total Weyl anomaly cancels to one loop\footnote{Although the concept of conformal anomaly  here is somewhat {\em  lato sensu}, in that it refers to the trace of the energy momentum tensor, even if the corresponding  classical theory is non-conformal.}. Moreover, given the absence of a non-renormalization theorem of the Adler-Bardeen type  for the Weyl anomaly \cite{Drummond}, it is not yet known whether or not this property can still hold to higher loops. It is clear that there is a lot of work ahead.

\section*{Acknowledgments}
We thank M.N. Chernodub, K. Landsteiner and M.A.H. Vozmediano for enjoyable as well as useful discussions. This work has received funding from the Spanish Research Agency (Agencia Estatal de Investigacion) through the grant IFT Centro de Excelencia Severo Ochoa SEV-2016-0597, and the European Union's Horizon 2020 research and innovation programme under the Marie Sklodowska-Curie grants agreement No 674896 and No 690575. We also have been partially supported by FPA2016-78645-P(Spain). RSG is supported by the Spanish FPU Grant No FPU16/01595. RSG is grateful to A. Herraez for useful comments on the manuscript.

\newpage

\newpage


\end{document}